\font\tenfrakturb=eufb10
\font\tenfraktur=eufm10
\font\tenmsbm=msbm10
\font\sevenfrakturb=eufb7
\font\sevenfraktur=eufm7
\font\sevenmsbm=msbm7
\font\fivefrakturb=eufb5
\font\fivefraktur=eufm5
\font\fivemsbm=msbm5
\newfam\bgothicfam
\newfam\gothicfam
\newfam\msbmfam
\textfont\bgothicfam = \tenfrakturb \scriptfont\bgothicfam=\sevenfrakturb
\scriptscriptfont\bgothicfam=\fivefrakturb
\textfont\gothicfam = \tenfraktur \scriptfont\gothicfam=\sevenfraktur
\scriptscriptfont\gothicfam=\fivefraktur
\textfont\msbmfam = \tenmsbm \scriptfont\msbmfam=\sevenmsbm
\scriptscriptfont\msbmfam=\fivemsbm

\def\Bbb{\tenmsbm\fam\msbmfam}

\catcode`@=11
\def\renewcounter#1{\@definecounter{#1}\@ifnextchar[{\@newctr{#1}}{}}

\documentstyle{elsart}
\begin{document}
\begin{flushleft}
Published in: {{\it Phys. Lett.} {\bf B617} (2005), No. 1--2 , pp. 67-77}
\end{flushleft}
\begin{frontmatter}
\title{Structure of the confining solutions for SU(3)-Yang-Mills equations
    and confinement mechanism }
\author{Yu. P. Goncharov}
\address{Theoretical Group, Experimental Physics Department, State Polytechnical
         University, Sankt-Petersburg 195251, Russia}

\begin{abstract} 
Structure of exact solutions modelling confinement is discussed for 
SU(3)-Yang-Mills equations and uniqueness of such confining solutions is proved 
in a certain sense. Relationship of the obtained results to QCD and the confinement 
mechanism is considered. Incidentally the Wilson confinement criterion for 
the found solutions is verified and also numerical estimates for 
strength of magnetic colour field responsible for linear confinement are 
adduced in the ground state of charmonium.
\end{abstract}
\begin{keyword}
Exact solutions \sep Quantum chromodynamics \sep Confinement
\PACS 12.38.-t \sep 12.38.Aw \sep 12.90.+b
\end{keyword}
\end{frontmatter}

\section{Introduction}

As was remarked in Ref. \cite{Gon01}, there exists a natural way of 
building meson spectroscopy and relativistic models of mesons which might be 
based on the exact solutions of the SU(3)-Yang-Mills equations modelling
quark confinement, the so-called confining solutions, and also on the 
corresponding modulo square integrable solutions of the Dirac equation in those  
confining SU(3)-fields. The given approach is the direct
consequence of the {\it relativistic} QCD (quantum chromodynamics) Lagrangian 
since the mentioned Yang-Mills and Dirac equations are derived just from the 
latter one. In Ref. \cite{Gon01} both the types of solutions were obtained and 
then in Ref. \cite{Gon03} they were successfully 
applied to the description of the charmonium spectrum. In its turn, the mentioned 
description pointed out the linear confinement to be (classically) governed by 
the magnetic colour field linear in $r$, the distance between quarks.

 The results of Refs. \cite{{Gon01},{Gon03}} suggested the following mechanism 
of confinement to occur within the framework of QCD (at any rate, for mesons
and quarkonia). The gluon exchange between quarks is realized in such a way
that at large distances it leads to the confining SU(3)-field which may be 
considered classically (the gluon concentration becomes huge and gluons form 
the boson condensate -- a classical field) and is 
a {\em nonperturbative} solution of the SU(3)-Yang-Mills 
equations. Under the circumstances mesons are the {\em relativistic bound 
states} described by the corresponding wave functions -- {\em nonperturbative} 
modulo square integrable solutions of the Dirac equation in this 
confining SU(3)-field. For each meson there 
exists its own set of real constants (for more details see Section 2) 
$a_j, A_j, b_j, B_j$ parametrizing the mentioned confining gluon
field (the gluon condensate) and the corresponding wave 
functions while the latter ones also depend on $\mu_0$, the reduced
mass of the current masses of quarks forming 
meson. It is clear that constants $a_j, A_j, b_j, B_j,\mu_0$
should be extracted from experimental data. Further application of the 
approach to quarkonia (charmonium and bottomonium) in 
Refs. \cite{{GC03},{GB04}} confirmed the mentioned picture of linear 
confinement and, in particular, gave possibility to estimate the above gluon 
concentrations.

The aim of the present Letter is to specify a number of features in the above 
confinement scenario, in particular, to show that the confining solutions 
found in Ref. \cite{Gon01} and used in Refs. \cite{{Gon03},{GC03},{GB04}} are 
in essence the unique ones.  

     Further we shall deal with the metric of 
the flat Minkowski spacetime $M$ that we write down 
[using the ordinary set of local spherical 
($r,\vartheta,\varphi$) 
or rectangular (Cartesian) ($x,y,z$) coordinates
for spatial part in the forms
$$ds^2=g_{\mu\nu}dx^\mu\otimes dx^\nu\equiv 
dt^2-dr^2-r^2(d\vartheta^2+\sin^2\vartheta d\varphi^2)\>, \eqno(1)$$
or 
$$ds^2=g_{\mu\nu}dx^\mu\otimes dx^\nu\equiv
dt^2-dx^2-dy^2-dz^2\>, \eqno(2)$$
so the components $g_{\mu\nu}$ take different values depending on the
choice of coordinates.
Besides we have $\delta=|\det(g_{\mu\nu})|=(r^2\sin\vartheta)^2$
in spherical coordinates and the exterior differential $d=\partial_t dt+
\partial_xdx+\partial_ydy+\partial_zdz$ or $d=\partial_t dt+\partial_r dr+
\partial_\vartheta d\vartheta+\partial_\varphi d\varphi$ in the corresponding
coordinates and we denote 3-dimensional vectors by bold font. 

  Throughout the paper we employ the system of units with $\hbar=c=1$,
unless explicitly stated otherwise. 

\section{Uniqueness of the confining solutions}
If $A=A_\mu dx^\mu=A^a_\mu \lambda_adx^\mu$ is a SU(3)-connection in the
(trivial) three-dimensional bundle $\xi$ over the Minkowski spacetime, 
where $\lambda_c$ are the known Gell-Mann matrices, then
we are interested in the confining solutions $A$ of the SU(3)-Yang-Mills 
equations
$$d\ast F= g(\ast F\wedge A - A\wedge\ast F) \>,\eqno(3)$$ 
while the curvature matrix (field strentgh)
for the $\xi$-bundle is 
$F=dA+gA\wedge A= F^a_{\mu\nu}\lambda_adx^\mu\wedge dx^\nu$ and $\ast$ 
means the Hodge star
operator conforming to metric (1), $g$ is a gauge coupling constant.

The confining solutions were defined in Ref. \cite{Gon01} as the 
spherically symmetric solutions of the Yang-Mills 
equations (3) containing only the components of the 
SU($3$)-field which are Coulomb-like or linear in $r$. Additionally 
we shall impose the Lorentz condition on the sought solutions. 
The latter condition is necessary for 
quantizing the gauge fields consistently within the framework of perturbation 
theory (see, e. g. Ref. \cite{Ryd85}), so we should impose the given condition 
that can be written
in the form ${\rm div}(A)=0$, where the divergence of the Lie algebra valued
1-form $A=A_\mu dx^\mu=A^a_\mu \lambda_adx^\mu$ is defined by the relation (see, e. g.
Refs. \cite{Bes87})
$${\rm div}(A)=\frac{1}{\sqrt{\delta}}\partial_\mu(\sqrt{\delta}g^{\mu\nu}
A_\nu)\>.\eqno(4)$$
It should be noted the following. If writing down the Yang-Mills equations (3) 
in components then we 
shall be drowned in a sea of indices which will strongly hamper searching for one 
or another ansatz and make it to be practically immense. Using the Hodge star 
operator as well as the rules of external calculus makes the problem to be 
quite foreseeable and quickly leads to the aim. Let us remind, therefore, the 
properties of Hodge star operator we shall need (for more details see 
Refs. \cite{Bes87}).
\subsection{Hodge star operator $*$ on Minkowski spacetime 
in spherical coordinates}
Let $M$ is a smooth manifold of dimension $n$ so we denote an 
algebra of smooth functions on $M$ as $F(M)$. In a standard way the spaces 
of smooth differential $p$-forms $\Lambda^p(M)$ ($0\le p\le n$) are defined 
over $M$ as modules over $F(M)$.  
If a (pseudo)riemannian metric $G=ds^2=g_{\mu\nu}dx^\mu\otimes dx^\nu$ is given 
on $M$ in local coordinates 
$x=(x^\mu)$ then $G$ can naturally be continued on spaces $\Lambda^p(M)$ 
by relation 
$$G(\alpha,\beta)={\rm det}\{G(\alpha_i,\beta_j)\}    \eqno(5)$$
for $\alpha=\alpha_1\wedge\alpha_2...\wedge\alpha_p$, 
$\beta=\beta_1\wedge\beta_2...\wedge\beta_p$, where for 1-forms
$\alpha_i=\alpha_\mu^{(i)}dx^\mu$, $\beta_j=\beta_\nu^{(j)}dx^\nu$ we have 
$G(\alpha_i,\beta_j)=g^{\mu\nu}\alpha_\mu^{(i)}\beta_\nu^{(j)}$ with the 
Cartan's wedge (external) product $\wedge$. Under the circumstances the Hodge 
star operator $*$: $\Lambda^p(M)\to\Lambda^{n-p}(M)$ is defined for any 
$\alpha\in\Lambda^p(M)$ by
$$\alpha\wedge(*\alpha)=G(\alpha,\alpha)\omega_g\> \eqno(6)$$
with the volume $n$-form 
$\omega_g=\sqrt{|\det(g_{\mu\nu})|}dx^1\wedge...dx^n$. 
For example, for 2-forms $F=F_{\mu\nu}dx^\mu\wedge dx^{\nu}$ we have
$$
F\wedge\ast F=(g^{\mu\alpha}g^{\nu\beta}-g^{\mu\beta}g^{\nu\alpha})
F_{\mu\nu}F_{\alpha\beta}
\sqrt{\delta}\,dx^1\wedge dx^2\cdots\wedge dx^n,\,\mu<\nu,\,\alpha<\beta 
\>\eqno(7)
$$
with $\delta=|\det(g_{\mu\nu})|$.
If $s$ is the number 
of (-1) in a canonical presentation of quadratic form $G$ then two most 
important properties of $*$ for us are 
$$ *^2=(-1)^{p(n-p)+s}\>,\eqno(8)$$
$$ *(f_1\alpha_1+f_2\alpha_2)=f_1(*\alpha_1)+f_2(*\alpha_2)\> \eqno(9)$$
for any $f_1, f_2 \in F(M)$, $\alpha_1, \alpha_2 \in\Lambda^p(M)$, i. e., 
$*$ is a $F(M)$-linear operator. Due to (9) for description 
of $*$-action in local coordinates it is enough to specify $*$-action on 
the basis elements of $\Lambda^p(M)$, i. e. on the forms 
$dx^{i_1}\wedge dx^{i_2}\wedge...\wedge dx^{i_p}$ with $i_1<i_2<...<i_p$ 
whose number is equal to $C_n^p=\frac{n!}{(n-p)!p!}$.

The most important case of $M$ in the given paper is the Minkowski spacetime 
with local coordinates $t, r, \vartheta, \varphi$, where 
$r, \vartheta, \varphi$ stand for spherical coordinates on spatial part of $M$. 
The metric is given by (1) and we shall obtain the $*$-action on the basis 
differential forms according to (6)
$$\ast dt=r^2\sin{\vartheta}dr\wedge d\vartheta\wedge d\varphi,\>
\ast dr=r^2\sin{\vartheta}dt\wedge d\vartheta\wedge d\varphi,\>$$
$$\ast d\vartheta=-r\sin{\vartheta}dt\wedge dr\wedge d\varphi,\>
\ast d\varphi=rdt\wedge dr\wedge d\vartheta,\>$$
$$\ast(dt\wedge dr)=-r^2\sin\vartheta d\vartheta\wedge d\varphi\>,
\ast(dt\wedge d\vartheta)=\sin\vartheta dr\wedge d\varphi\>,$$
$$\ast(dt\wedge d\varphi)=-\frac{1}{\sin\vartheta}dr\wedge d\vartheta\>,
\ast(dr\wedge d\vartheta)=\sin\vartheta dt\wedge d\varphi\>,$$
$$\ast(dr\wedge d\varphi)=-\frac{1}{\sin\vartheta}dt\wedge d\vartheta\>,
\ast(d\vartheta\wedge d\varphi)=\frac{1}{r^2\sin\vartheta}dt\wedge dr\>,$$
$$\ast(dt\wedge dr\wedge d\vartheta)=\frac{1}{r}d\varphi\>,
\ast(dt\wedge dr\wedge d\varphi)=-\frac{1}{r\sin{\vartheta}}d\vartheta,\>$$
$$\ast(dt\wedge d\vartheta\wedge d\varphi)=\frac{1}{r^2\sin{\vartheta}}dr,\>
\ast(dr\wedge d\vartheta\wedge d\varphi)=
\frac{1}{r^2\sin{\vartheta}}dt,\>
\eqno(10)$$
so that on 2-forms $\ast^2=-1$, as should be in accordance with 
(8).

At last it should be noted that all the above is easily over linearity 
continued on the matrix-valued differential forms (see, e. g., 
Ref. \cite{Car}), i. e., on the arbitrary 
linear combinations of forms 
$a_{i_1i_2...i_p}dx^{i_1}\wedge dx^{i_2}\wedge...\wedge dx^{i_p}$, 
where coefficients $a_{i_1i_2...i_p}$ belong to some space of matrices $V$, 
for example, a SU($3$)-Lie algebra. But now the 
Cartan's wedge (external) product $\wedge$ should be understood as  
product of matrices with elements consisting of usual (scalar) differential 
forms. In the SU(3)-case, if $T_a$ are 
matrices of 
generators of the SU($3$)-Lie algebra in $3$-dimensional representation, we 
continue the above scalar product $G$ on the SU($3$)-Lie algebra valued 1-forms 
$A=A^a_\mu T_adx^\mu$ and $B=B^b_\nu T_bdx^\nu$ by the relation
$$G(A,B)=g^{\mu\nu}A^a_\mu B^b_\nu{\rm Tr}(T_aT_b)\,,\eqno(11)$$
where Tr signifies the trace of a matrix, 
and, on linearity with the help of (5), $G$ can be continued 
over any SU($3$)-Lie algebra valued forms.
\subsection{The confining solutions}
Let us for definiteness put $T_a=\lambda_a$. 
Under this situation we can take the general ansatz of form 
$$A=r^\mu\mit\Gamma dt+A_rdr+A_\vartheta d\vartheta+
r^\nu\mit\Delta d\varphi \>,\eqno(12)$$
where $A_\vartheta= r^\rho T$ and matrices $\mit\Gamma=\alpha^a\lambda_a$, 
$\mit\Delta=\beta^a\lambda_a$, $T=\gamma^a\lambda_a$, $A_r=f^a(r)\lambda_a$ 
with arbitrary real 
constants $\alpha^a, \beta^a, \gamma^a$ and arbitrary real functions $f^a(r)$. 
It could seem that 
there is a more general ansatz in the form 
$A=r^{\mu_a}\alpha^a\lambda_a dt+f^a(r)\lambda_adr+
r^{\rho_a}\gamma^a\lambda_a d\vartheta+
r^{\nu_a}\beta^a\lambda_a d\varphi$ but somewhat more complicated considerations 
than the ones below show that all the same we should have 
$\mu_a=\mu, \nu_a=\nu, \rho_a=\rho$ for any $a$ so we at once consider this 
condition to be fulfilled to avoid unnecessary complications. 
Then for the above ansatz (12) the Lorentz condition ${\rm div}(A)=0$ 
takes the form
$$\partial_r(r^2\sin{\vartheta}g^{rr}A_r)+
\partial_\vartheta(r^2\sin{\vartheta}g^{\vartheta\vartheta}A_\vartheta)=0$$
which can be rewritten as 
$$\partial_\vartheta(\sin{\vartheta}A_\vartheta)+\sin{\vartheta}
\partial_r(r^2A_r)=0\>, \eqno(13)$$
wherefrom it follows $\cot{\vartheta}r^\rho T+\partial_r(r^2A_r)=0$ while 
the latter entails 
$$A_r=\frac{C}{r^2}-\frac{\cot{\vartheta}r^{\rho-1}T}{\rho+1}\>
\eqno(14)$$
with a constant matrix $C$.
Then we can see that it should put $C=T=0$ or else $A_r$ will not be spherically 
symmetric and the confining one where only the powers of $r$ equal to $\pm1$ 
are admissible. As a result we come to the conclusion that one should put 
$A_r=A_\vartheta=0$ in (12).
After this we have ($[,]$ signifies matrix commutator) 
$F=dA+gA\wedge A=-\mu r^{\mu-1}\mit\Gamma dt\wedge dr+ 
\nu r^{\nu-1}\mit\Delta dr\wedge d\varphi+gr^{\mu+\nu}
[\mit\Gamma,\mit\Delta]dt\wedge d\varphi$ which entails [with the help of 
(10)]   
$$\ast F=\mu r^{\mu+1}\sin{\vartheta}\mit\Gamma d\vartheta\wedge d\varphi-
\frac{\nu r^{\nu-1}\mit\Delta}{\sin{\vartheta}} dt\wedge d\vartheta-
\frac{gr^{\mu+\nu}}{\sin{\vartheta}}
[\mit\Gamma,\mit\Delta]dr\wedge d\vartheta$$ 
and the Yang-Mills equations (3) turn into
$$\mu(\mu+1)r^{\mu}\sin^2{\vartheta}\mit\Gamma=
g^2r^{\mu+2\nu}[\mit\Delta,[\mit\Gamma,\mit\Delta]]\>,$$
$$\nu(\nu-1)r^{\nu-2}\mit\Delta=
g^2r^{2\mu+\nu}[\mit\Gamma,[\mit\Gamma,\mit\Delta]]\>.
\eqno(15)$$
It is now not complicated to enumerate possibilities for obtaining the confining 
solutions in accordance with (15), where we should put $\mu=-1$, $\nu=1$.
\begin{enumerate}
\item $\mit\Gamma$=0 or $\mit\Delta$=0. This situation does obviously not 
correspond to a confining solution
\item $\mit\Gamma=C_0\mit\Delta$ with some constant $C_0$. This case conforms 
to that all the parameters $\alpha^a$ describing electric colour Coulomb field 
(see Subsection 2.3) and the ones $\beta^a$ for linear magnetic colour 
field are 
proportional -- the situation is not quite clear from physical point of view
\item Matrices $\mit\Gamma,\mit\Delta$ are not equal to zero simultaneously and 
both matrices belong to Cartan subalgebra of SU(3)-Lie algebra. The parameters 
$\alpha^a, \beta^a$ of electric and magnetic colour fields are not connected 
and arbitrary, i. e. 
they should be chosen from experimental data. The given situation is the most 
adequate to the physics in question and the corresponding confining solution 
is in essence the same which has been obtained as far back as in 
Ref. \cite{Gon01} from other considerations. 
\end{enumerate}
One can slightly generalize the starting ansatz (12) taking it in the form 
$A=(r^\mu\mit\Gamma +A')dt+(r^\nu\mit\Delta +B')d\varphi$ 
with matrices $A'=A^a\lambda_a$, $B'=B^a\lambda_a$ and constants $A^a,B^a$. 
Then considerations along the same above lines draw the conclusion that 
the nontrivial confining solution is described by 
$\mit\Gamma,\mit\Delta, A', B'$ belonging to Cartan subalgebra. 

Let us remind that, by definition, 
a Cartan subalgebra is a maximal abelian subalgebra in 
the corresponding Lie algebra, i. e., the commutator for any two matrices of 
the Cartan subalgebra is equal to zero (see, e.g., Ref. \cite{Bar}). For 
SU(3)-Lie algebra the conforming Cartan subalgebra is generated by the 
Gell-Mann matrices $\lambda_3, \lambda_8$ which are
$$\lambda_3=\pmatrix{1&0&0\cr 0&-1&0\cr 0&0&0\cr}\,,  
  \lambda_8={1\over\sqrt3}\pmatrix{1&0&0\cr 0&1&0\cr 
                   0&0&-2\cr}\,.\eqno(16)$$
Then it is easy to gain the only nontrivial 
solution in question in the form (which reflects the fact that for any matrix 
${\cal T}$ from SU(3)-Lie algebra we have ${\rm Tr}\,{\cal T}=0$) 
 $$ A^3_t+\frac{1}{\sqrt{3}}A^8_t =-\frac{a_1}{r}+A_1 \>,
 -A^3_t+\frac{1}{\sqrt{3}}A^8_t=-\frac{a_2}{r}+A_2\>,
-\frac{2}{\sqrt{3}}A^8_t=\frac{a_1+a_2}{r}-(A_1+A_2)\>, $$
$$ A^3_\varphi+\frac{1}{\sqrt{3}}A^8_\varphi =b_1r+B_1 \>,
 -A^3_\varphi+\frac{1}{\sqrt{3}}A^8_\varphi=b_2r+B_2\>,
-\frac{2}{\sqrt{3}}A^8_\varphi=-(b_1+b_2)r-(B_1+B_2)\>, \eqno(17)$$
where real constants $a_j, A_j, b_j, B_j$
parametrize the solution, and we wrote down
the solution in the combinations that are just
needed to insert into the corresponding Dirac 
equation (see Section 3). From here it follows one more form 
$$A^3_t = [(a_2-a_1)/r+A_1-A_2]/2,\>
A^8_t =[A_1+A_2-(a_1+a_2)/r]\sqrt{3}/2\>,$$
$$ A^3_\varphi = [(b_1-b_2)r+B_1-B_2]/2,
   A^8_\varphi= [(b_1+b_2)r+B_1+B_2]\sqrt{3}/2\>\eqno(18)$$

Clearly, the obtained results may be extended over all SU($N$)-groups with 
$N\ge2$ and even 
over all semisimple compact Lie groups since for them the corresponding Lie 
algebras possess just the only Cartan subalgebra. Also we can talk about the 
compact non-semisimple groups, for example, U($N$). In the latter case 
additionally to Cartan subalgebra we have centrum consisting from the matrices 
of the form $\alpha I_N$ ($I_N$ is the unit matrix $N\times N$) with arbitrary 
constant $\alpha$. 
The most relevant physical cases are of course U(1)- and SU(3)-ones 
(QED and QCD), 
therefore we shall not consider further generalizations of the results 
obtained but let us write out the corresponding solution for U(1)-case which 
will useful for to interpret above solutions (17)--(18) in the more habitual 
physical terms. It should also be noted that the 
nontrivial confining solutions obtained exist at any gauge coupling constant 
$g$, i. e. they are essentially {\em nonperturbative} ones.
\subsection{\rm{U(1)}-case}
Under this situation the Yang-Mills equations (3) turn into the second pair of 
Maxwell equations 
$$d\ast F= 0 \eqno(19)$$ 
with $F=dA$, $A=A_\mu dx^\mu$. 
We search for 
the solution of (19) in the form $A=A_t(r)dt+A_\varphi(r)d\varphi$.
It is then easy to check that $F=dA=
-\partial_rA_tdt\wedge dr+\partial_rA_\varphi dr\wedge d\varphi$ and
according to (10) we get
$\ast F=r^2\sin\vartheta\partial_rA_td\vartheta\wedge d\varphi-
\frac{1}{\sin\vartheta}\partial_rA_\varphi dt\wedge d\vartheta$. From here
it follows that (19) yields
$$\partial_r(r^2\partial_rA_t)=0,\>\partial^2_rA_\varphi=0\>,\eqno(20)$$
and we write down the solutions of (20) as 
$$ A_t =\frac{a}{r}+A \>, A_\varphi=br+B \>\eqno(21)$$
with some constants $a, b, A, B$ parametrizing solutions (further for the 
sake of simplicity let us put $a=1, b=1\ {\rm GeV}, A=B=0$).
 
To interpret solutions (21) in the more habitual physical terms let us pass on
to Cartesian coordinates employing the relations 
$$\varphi=\arctan(y/x),\>
d\varphi=\frac{\partial\varphi}{\partial x}dx+
\frac{\partial\varphi}{\partial y}dy\> \eqno(22)$$
which entails
$$A_\varphi d\varphi=-\frac{ry}{x^2+y^2}dx+\frac{rx}{x^2+y^2}dy\>\eqno(23)$$
and we conclude that the solutions of (21) describe the combination
of the electric Coulomb field with potential $\Phi=A_t$ and the constant 
magnetic field with vector-potential
$${\bf A}=(A_x,A_y,A_z)=(-\frac{ry}{x^2+y^2},\frac{rx}{x^2+y^2},0)=
(-\frac{\sin\varphi}{\sin\vartheta},\frac{\cos\varphi}{\sin\vartheta},0)\>,
\eqno(24)$$
which is {\it linear} in $r$ in spherical coordinates and the 3-dimensional 
divergence ${\rm div}{\bf A}=0$, as can be checked directly. Then Eqs. (19) in 
Cartesian coordinates takes the form
$$\Delta\Phi=0,\> {\rm rotrot}{\bf A}= \Delta{\bf A}=0\> \eqno(25)$$
with the Laplace operator $\Delta=\partial_x^2+\partial_y^2+\partial_z^2$.
At last, it is easy to check that the solution under consideration satisfies the 
Lorentz condition ${\rm div}(A)=0$.

Practically the same considerations as the above ones in electrodynamics
show that the solutions (17)--(18) describe the configuration of the electric 
Coulomb-like colour field (components $A_t$) with potentials $\Phi^3,\Phi^8$ 
and the constant magnetic colour 
field (components $A_\varphi$) with vector-potentials ${\bf A}^3,{\bf A}^8$
which are {\it linear} in $r$ in spherical coordinates with 3-dimensional 
divergences ${\rm div}{\bf A^{3,8}}=0$ and 
$\Delta\Phi^{3,8}=0,\> {\rm rotrot}{\bf A^{3,8}}= \Delta{\bf A^{3,8}}=0$.  

\section{Relationship with QCD}
The previous section leads us to the following problem: how to describe 
possible relativistic bound states in the obtained confining SU(3)-Yang-Mills 
fields? 
The sought description should be obviously based on the QCD-Lagrangian. 
Let us write down this Lagrangian (for one flavour) in 
arbitrary curvilinear (local) coordinates in Minkowski spacetime 
$${\cal L}=\overline{\Psi}{\cal D}\Psi -\mu_0\overline{\Psi}\Psi 
-\frac{1}{4}(g^{\mu\alpha}g^{\nu\beta}-g^{\mu\beta}g^{\nu\alpha})
F^a_{\mu\nu}F^a_{\alpha\beta}, \,\mu<\nu,\,\alpha<\beta
\eqno(26)$$
where, if denoting
$S(M)$ and $\xi$, respectively, the standard spinor bundle
and $3$-dimensional vector one (equipped with a SU($3$)-connection with 
the corresponding connection and curvature matrices 
$A=A_\mu dx^\mu=A^a_\mu \lambda_adx^\mu$, 
$F=dA+gA\wedge A=F^a_{\mu\nu}\lambda_adx^\mu\wedge dx^\nu$) 
over Minkowski spacetime, we can
construct tensorial product $\Xi=S(M)\otimes\xi$. It is clear that $\Psi$ is 
just a section of the latter bundle, i. e. 
$\Psi$ can be chosen in the form
$\Psi=(\Psi_1,\Psi_2, \Psi_3)$
with the four-dimensional Dirac spinors $\Psi_j$ representing the $j$-th colour
component while $\overline{\Psi}=\Psi^{\dag}(\gamma^0\otimes I_3)$ is 
the adjont spinor,($\dag$) stands for hermitian conjugation, 
$\otimes$ means tensorial product of matrices,  
$\mu_0$ is a mass parameter, ${\cal D}$ is the Dirac operator
with coefficients in $\xi$ (see below). At last, we have  
the condition ${\rm Tr}(\lambda_a\lambda_b)=2\delta_{ab}$ so that the 
third addendum in (26) has the form $G(F,F)/8$ with $G$ of (11), where 
coefficient $1/8$ is chosen from physical considerations.

From general considerations (see, e. g., Ref. \cite{89}) the explicit form of
the operator ${\cal D}$ in local coordinates $x^\mu$ on Minkowski spacetime 
can be written as follows
$${\cal D}=i(\gamma^e\otimes I_3)E_e^\mu\left(\partial_\mu\otimes I_3
-\frac{1}{2}\omega_{\mu ab}\gamma^a\gamma^b\otimes I_3-igA_\mu\right),
\>a < b ,\>\eqno(27)$$
where the forms\ $\omega_{ab}=\omega_{\mu ab}dx^\mu$ obey 
the Cartan structure equations
$de^a=\omega^a_{\ b}\wedge e^b$, while the
orthonormal basis $e^a=e^a_\mu dx^\mu$ in cotangent bundle and
dual basis $E_a=E^\mu_a\partial_\mu$ in tangent bundle are connected by the
relations $e^a(E_b)=\delta^a_b$. At last, matrices $\gamma^a$ represent
the Clifford algebra of
the quadratic form $Q_{1,3}=x_0^2-x_1^2-x_2^2-x_3^2$ in ${\Bbb C}^{4}$. 
It should be noted that Greek indices $\mu,\nu,...$
are raised and lowered with $g_{\mu\nu}$ of (1) or its inverse $g^{\mu\nu}$
and Latin indices $a,b,...$ are raised and lowered by
$\eta_{ab}=\eta^{ab}$= diag(1,-1,-1,-1) except for Latin indices connected 
with SU($3$)-Lie algebras, 
so that $e^a_\mu e^b_\nu g^{\mu\nu}=\eta^{ab}$,
$E^\mu_aE^\nu_bg_{\mu\nu}=\eta_{ab}$ and so on but $\lambda_a=\lambda^a$.

Under the circumstances we can obtain the following equations according to the 
standard prescription of Lagrange approach from (26)
$${\cal D}\Psi=\mu_0\Psi\>,\eqno(28)$$
$$d\ast F= g(\ast F\wedge A - A\wedge\ast F) +gJ\>,\eqno(29)$$
where the source $J$ (a nonabelian SU($3$)-current) is 
$$J=j_\mu^a\lambda_a\ast(dx^\mu)=\ast j=\ast(j_\mu^a\lambda_adx^\mu)=
\ast(j^a\lambda_a)\>\eqno(30)$$
with currents 
$$j^a=j_\mu^adx^\mu=
\overline{\Psi}(\gamma_\mu\otimes I_3)\lambda^a\Psi\,dx^\mu\>,$$
so summing over 
$a=1,...,8$ is implied in (26) and (30). 

When using the relation (see, e. g. Refs. \cite{Fin}) 
$\gamma^cE_e^\mu\omega_{\mu ab}\gamma^a\gamma^b=
\omega_{\mu ab}\gamma^\mu\gamma^a\gamma^b=
-{\rm div}(\gamma)$ with matrix 1-form $\gamma=\gamma_\mu dx^\mu$ 
[where ${\rm div}$ is defined by relation (4)] and also the 
fact that $(\gamma^\mu)^{\dag}\gamma^0=\gamma^0\gamma^\mu$, the 
Dirac equation for spinor $\overline{\Psi}$ will be  
$$i\partial_\mu\overline{\Psi}(\gamma^\mu\otimes I_3)+
\frac{i}{2}\overline{\Psi}{\rm div}(\gamma)\otimes I_3-
g\overline{\Psi}(\gamma^\mu\otimes I_3)A^a_\mu \lambda_a=-\mu_0\overline{\Psi}
\>.\eqno(28')$$
Then multiplying (28) by $\overline{\Psi}\lambda_a$ from left and ($28'$) by 
$\lambda_a\Psi$ from right and adding the obtained equations, we get 
${\rm div}(j^a)={\rm div}(j)=0$ if spinor ${\Psi}$ obeys the Dirac equation 
(28).

The question now is how to connect the sought relativistic bound states with 
the system (28)--(29). To understand it let us apply to the experience 
related with QED. In the latter case Lagrangian looks like (26) with changing 
group SU($3$)$\to$U(1) so $\Psi$ will be just a four-dimensional Dirac spinor. 
Then, as is known (see, e. g. Ref. \cite{LL1}), when passing on to the 
nonrelativistic limit the Dirac equation (28) converts into the Pauli equation 
and further, if neglecting the particle spin, into the Schr{\"o}dinger 
equation, parameter $\mu_0$ becoming the reduced mass of two-body system. 
The modulo square integrable solutions of the Schr{\"o}dinger equation just 
describe bound states of a particle with mass $\mu_0$, or, that is equivalent, 
of the corresponding two-body system. Historically, however, everything was 
just vice versa. At first there appeared the Schr{\"o}dinger equation, then 
the Pauli and Dirac ones and only then the QED Lagrangian. In its turn, 
possibility of writing two-body Schr{\"o}dinger equation on the whole owed 
to the fact that the corresponding two-body problem in classical 
nonrelativistic (newtonian) mechanics was well posed and actually quantizing 
the latter gave two-body Schr{\"o}dinger equation. Another matter was Dirac 
equation. Up to now nobody can say what two-body problem in classical 
relativistic (einsteinian) mechanics could correspond to Dirac equation. The 
fact is that the two-body problem in classical relativistic mechanics has 
so far no single-valued statement. Conventionally, therefore, Dirac 
equation in QED is treated as the relativistic wave equation describing one 
particle with spin one half in an external electromagnetic field. 

There is, however, one important exclusion -- the hydrogen atom. When solving 
the Dirac equation here one considers mass parameter $\mu_0$ to be equal to the 
electron mass and one gets the so-called Sommerfeld formula for hydrogen atom 
levels which passes on to the standard Schr{\"o}dinger formula for hydrogen atom 
spectrum in nonrelativistic limit (for more details see, e. g. 
Ref. \cite{LL1}). But in the Schr{\"o}dinger formula 
mass parameter $\mu_0$ is equal to the reduced mass of electron and proton. As 
a consequence, 
it is tacitly supposed that in Dirac equation the mass parameter should be 
equal to the same reduced mass of electron and proton as in Schr{\"o}dinger 
equation. Just the mentioned reduced mass is approximately equal to that of 
electron but, exactly speaking, it is not the case. We remind that for the 
problem under discussion (hydrogen atom) the external field is the Coulomb 
electric one between electron and proton, essentially nonrelativistic object 
in the sense that it does not vanish in nonrelativistic limit at $c\to\infty$.
If now to place hydrogen atom in a magnetic field then obviously spectrum of 
bound states will also depend on parameters decribing the magnetic field. 
The latter, however, is essentially relativistic object and vanishes at 
$c\to\infty$ because, as is well known, in the world with $c=\infty$ there 
exist no magnetic fields (see any elementary textbook on physics, e. g. 
Ref. \cite{Sav82}). But it is clear that spectrum should as before depend 
on $\mu_0$ as well and we can see that $\mu_0$ is the same reduced mass as 
before since in nonrelativistic limit we again should come to the hydrogen 
atom spectrum with the reduced mass. So we can draw the conclusion that if
an electromagnetic field is a combination of electric Coulomb field between two 
charged elementary particles and some magnetic field (which may be generated 
by the particles themselves) then there are certain grounds to consider the 
given (quantum) two-body problem to be equivalent to the one of motion for one 
particle with usual reduced mass in the mentioned electromagnetic field. As a 
result, we can use the Dirac equation for finding possible relativistic 
bound states for such a particle implying that this is really some description 
of the corresponding two-body problem. 

Actually in QED the situaion is just as the described one but magnetic field 
is usually weak and one may restrict 
oneself to some corrections from this field to the nonrelativistic Coulomb 
spectrum (e. g., in the Seemann effect). If the magnetic field is strong then 
one should solve just Dirac 
equation in a nonperturbative way (see, e.g. Ref. \cite{ST}). The latter 
situation seems to be natural in QCD where the corresponding (colour) magnetic 
field should be very strong (see Section 4) 
because just it provides linear confinement of quarks as we could see above 
(see also Refs. \cite{{Gon03},{GC03},{GB04}}) and the given field 
also vanishes in nonrelativistic limit (for more details see 
Refs. \cite{{Gon03},{GC03}}). 

At last, we should make an important point that in QED the mentioned 
electromagnetic field is by definition always a solution of the Maxwell 
equations so 
within QCD we should require the confining SU(3)-field to be a solution of 
Yang-Mills equations. Consequently, returning to the system (28)--(29), we 
can suggest to decribe relativistic bound states of two quarks (mesons) in QCD 
by the compatible solutions of the given system. To be more precise, the meson 
wave functions should be the {\em nonperturbative} modulo square integrable 
solutions of Dirac equation (28) (with the above reduced mass $\mu_0$) in the 
confining SU(3)-Yang-Mills field being a {\em nonperturbative} solution of 
(29). In general case, however, the analysis of (29) is difficult because of 
availability of the nonabelian current $J$ of (30) in the right-hand side of 
(29) but we may use the circumstance that the corresponding modulo square 
integrable solutions of Dirac equation (28) might consist from the components 
of form $\Psi_j\sim r^{\alpha_j}e^{-\beta_j r}$ with some $\alpha_j>0$, 
$\beta_j>0$ 
which entails all the components of the current $J$ to be modulo $<< 1$ at each 
point of Minkowski space. The latter will allow us to put $J\approx0$ and we 
shall come to the problem of finding the confining solutions for the Yang-Mills 
equations of (29) with $J=0$ whose unique nontrivial form has been described in 
previous Section and after inserting the 
found solutions into Dirac equation (28) we should require the corresponding 
solutions of Dirac equation to have the above necessary behaviour. Under 
the circumstances the problem becomes self-consistent and can be analyzable 
and this has been done actually in Ref. \cite{Gon01}. 

It is clear that all the above considerations can be justified only by  
comparison with experimental data but now we obtain some intelligible 
programme of further activity which has been partly realized in 
Refs. \cite{{Gon03},{GC03},{GB04}}. 
\section{Verification of confinement criterion and estimates of colour magnetic 
field strength}
To illustrate some of the above let us verify the Wilson confinement criterion 
\cite{Wil} for the confining solutions (17)--(18) and also adduce numerical 
estimates for strength of colour magnetic field responsible for linear 
confinement in the ground state of charmonium. 
\subsection{Verification of confinement criterion} 
The Wilson confinement criterion is in essence the assertion that the so-called 
Wilson loop $W(c)$ should be subject to the area law for the confining gluonic 
field configuration. In its turn, the latter law is equivalent to the fact that 
energy $E(R)$ of the mentioned configuration (gluon condensate) is linearly 
increasing with $R$, a characteristic size of some volume $V$ containing the 
condensate. We can easily evaluate $E(R)$ for solutions (17)--(18) using the 
$T_{00}$-component (volumetric energy density) of 
the energy-momentum tensor for a SU($3$)-Yang-Mills field
$$T_{\mu\nu}={1\over4\pi}\left(-F^a_{\mu\alpha}\,F^a_{\nu\beta}\,g^{\alpha\beta}+
{1\over4}F^a_{\beta\gamma}\,F^a_{\alpha\delta}g^{\alpha\beta}g^{\gamma\delta}
g_{\mu\nu}\right)\>. \eqno(31) $$
It is not complicated to obtain the curvature matrix (field strentgh) 
corresponding to the solution (17)--(18) 
$$F= F^a_{\mu\nu}\lambda_a dx^\mu\wedge dx^\nu=
-\partial_r(A^a_t\lambda_a)dt\wedge dr
+\partial_r(A^a_\varphi \lambda_a)dr\wedge d\varphi
\>, \eqno(32)$$
which entails the only nonzero components
$$F^3_{tr}=\frac{a_1-a_2}{2r^2},\>
F^8_{tr}=\frac{(a_1+a_2)\sqrt{3}}{2r^2},\>
F^3_{r\varphi}=\frac{b_1-b_2}{2},\>
F^8_{r\varphi}=\frac{(b_1+b_2)\sqrt{3}}{2}\>\eqno(33) $$
and, in its turn,
$$T_{00}\equiv T_{tt}=\frac{1}{4\pi}\left\{\frac{3}{4}\left[(F^3_{tr})^2+
(F^8_{tr})^2\right]
+\frac{1}{4r^2\sin^2{\vartheta}}\left[(F^3_{r\varphi})^2+
(F^8_{r\varphi})^2\right]\right\}= $$
$$\frac{3}{16\pi}\left(\frac{a_1^2+a_1a_2+a_2^2}{r^4}+
\frac{b_1^2+b_1b_2+b_2^2}{3r^2\sin^2{\vartheta}}\right)\equiv
\frac{{\cal A}}{r^4}+
\frac{{\cal B}}{r^2\sin^2{\vartheta}}\>\eqno(34)$$
with ${\cal A}>0$, ${\cal B}>0$.
Let $V$ be the volume between two concentric spheres with radii $R_0<R$. Then 
$E(R)=\int_VT_{00}\sqrt{\delta}d^3x=
\int_VT_{00}r^2\sin{\vartheta}dr d\vartheta d\varphi\>$. As was discussed in 
Ref.\cite{{GB04}}, the notion of classical gluonic field is applicable only 
at distances $>> \lambda_B$, the de Broglie wavelength of quark, so we may  
take $R_0$ to be of order of a characteristic size of hadron (meson). Besides it 
should be noted that classical $T_{00}$ of (34) has a singularity along 
$z$-axis ($\vartheta=0,\pi$) and we have to introduce some angle $\vartheta_0$ 
whose physical meaning is to be clarified a little below. Under the circumstances, with 
employing the relations $\int d\vartheta/\sin{\vartheta}=\ln\tan{\vartheta/2}$, 
$\tan{\vartheta/2}=\sin{\vartheta}/(1+ \cos{\vartheta})=
(1-\cos{\vartheta})/\sin{\vartheta}$, we shall have 
$$E(R)=\int_{R_0}^R\int^{\pi-\vartheta_0}_{\vartheta_0}\int_{0}^{2\pi}
\left(\frac{{\cal A}}{r^2}+\frac{{\cal B}}{\sin^2{\vartheta}}\right)
\sin{\vartheta}dr d\vartheta d\varphi=E_0-\frac{4\pi{\cal A}}{R}+
{2\pi\cal B}R\ln\frac{1+\cos{\vartheta_0}}{1-\cos{\vartheta_0}} \>\eqno(35)$$
with $E_0=\frac{{4\pi\cal A}}{R_0}-
{2\pi\cal B}R_0\ln\frac{1+\cos{\vartheta_0}}{1-\cos{\vartheta_0}}$. It is clear 
that at $R\to\infty$ we get $E(R)\sim R$, i. e. the area law is fulfilled. 
\subsection{Estimates of colour magnetic field strength}
To estimate $R_0$ and ${\vartheta_0}$ under real physical situation let us 
consider the ground state of charmonium $\eta_c(1S)$ and apply the ideology of 
Section 3 for describing the given state. We use the parametrization of 
relativistic spectrum of charmonium from Ref. \cite{GB04}. Namely, Table 1 
contains necessary parameters of solutions (17)--(18) for the sought estimates.
\begin{table}[htbp]
\caption{Gauge coupling constant, mass parameter $\mu_0$ and
parameters of the confining SU(3)-gluonic field for charmonium.}
\label{t.1}
\begin{center}
\begin{tabular}{|c|c|c|c|c|c|c|c|}
\hline
\small $ g$ & \small $\mu_0$ & \small $a_1$  & \small $a_2$ & \small $b_1$ 
& \small $b_2$ & \small $B_1$ & \small $B_2$ \\
  & \small (GeV) &  &  & \small (GeV) & \small (GeV) & &  \\
\hline
\small 0.46900 & \small 0.62500 & \small 2.21104 & \small -0.751317 
& \small 20.2395 & \small -12.6317 & \small 6.89659 & \small  6.89659 \\
\hline
\end{tabular}
\end{center}
\end{table}
As for parameters $A_{1,2}$ of (17)--(18) then we can put them equal to zero 
\cite{GB04}. 
One can note that the reduced mass parameter $\mu_0$ is consistent with the
present-day experimental limits \cite{pdg} where the current mass of $c$-quark
($2\mu_0$) is accepted between 1.1 GeV and 1.4 GeV. Then wave function 
$\Psi=(\Psi_1,\Psi_2, \Psi_3)$ of the 
state under consideration [as a modulo square integrable solution of Dirac 
equation (28) in the field (17)--(18)] is given by 
the form (with Pauli matrix $\sigma_1$)
$$\Psi_j=e^{i\omega_j t}r^{-1}\pmatrix{F_{j1}(r)\Phi_j(\vartheta,\varphi)\cr\
F_{j2}(r)\sigma_1\Phi_j(\vartheta,\varphi)}\>,j=1,2,3\eqno(36)$$
with the 2D eigenspinor $\Phi_j=\pmatrix{\Phi_{j1}\cr\Phi_{j2}}$ of the
euclidean Dirac operator on the unit sphere ${\Bbb S}^2$.
The explicit form of $\Phi_j$ is not needed here and
can be found in Ref. \cite{GB04}. The radial parts are
$$F_{j1}=C_jP_jr^{\alpha_j}e^{-\beta_jr}\left(1-
\frac{gb_j}{\beta_j}\right),F_{j2}=iC_jQ_jr^{\alpha_j}e^{-\beta_jr}\left(1+
\frac{gb_j}{\beta_j}\right)\eqno(37)$$
with $\alpha_j=\sqrt{\Lambda_j^2-g^2a_j^2}$, $\Lambda_j=-1-gB_j$, 
$\beta_j=\sqrt{\mu_0^2-\omega_j^2+g^2b_j^2}$, $P_j=gb_j+\beta_j$, 
$Q_j=\mu_0+\omega_j$ where $a_3=-(a_1+a_2),b_3=-(b_1+b_2), B_3=-(B_1+B_2)$ and 
constants $C_j$ are determined
from the normalization conditions 
$\int_0^\infty(|F_{j1}|^2+|F_{j2}|^2)dr=1/3$. Energy of the ground state is 
obtained as 
$$\epsilon=\sum\limits_{j=1}^3\omega_j\equiv
\frac{-\Lambda_1g^2a_1b_1+\alpha_1|\Lambda_1|\mu_0}{\Lambda_1^2}+
\frac{-\Lambda_2g^2a_2b_2+\alpha_2|\Lambda_2|\mu_0}{\Lambda_2^2}+
\frac{-\Lambda_3g^2a_3b_3-\alpha_3|\Lambda_3|\mu_0}{\Lambda_3^2}=
2.9796\ {\rm GeV}\>.
\eqno(38)$$
Then we can put $R_0=\frac{1}{3}\sum_{j=1}^{3}1/\beta_j
\approx0.0399766\ {\rm fm}$ and, adding the rest energies of quarks $2(2\mu_0)$ 
to $E_0$ of (35), we consider $\epsilon=E_0+4\mu_0=2.9796\ {\rm GeV}$ which 
entails $\vartheta_0\approx0.723388$ and one may consider $\vartheta_0$ to be 
the parameter determining some cone $0\le\vartheta\le\vartheta_0$ so the quark  
emits gluons outside of the cone. It is evident that the Wilson criterion is 
purely classic one: quantum mechanically the stay probability of quarks at 
distances $>R_0$ is of order $\sum_{j=1}^{3}(|F_{j1}|^2+|F_{j2}|^2)$ and 
is damped exponentially fast at $R>R_0$. Just the colour magnetic field defines 
this damping through the coefficients $\beta_j$. 

At last, using the Hodge star operator in 3-dimensional euclidean space [where
$ds^2=g_{\mu\nu}dx^\mu\otimes dx^\nu\equiv 
dr^2+r^2(d\vartheta^2+\sin^2\vartheta d\varphi^2)$, 
$\ast(dr\wedge d\vartheta)=\sin{\vartheta}d\varphi, 
\ast(dr\wedge d\varphi)=-d\vartheta/\sin{\vartheta},
\ast(d\vartheta\wedge d\varphi)=dr/(r^2\sin{\vartheta}$],
we can confront components $F^3_{r\varphi},F^8_{r\varphi}$ of (33) 
with 3-dimensional 1-forms of the magnetic colour field 
$${\bf H}^3=-\frac{b_1-b_2}{2\sin{\vartheta}}d\vartheta\>,
{\bf H}^8=-\frac{(b_1+b_2)\sqrt{3}}{2\sin{\vartheta}}d\vartheta $$ 
that are modulo equal to 
$\sqrt{g^{\mu\nu}{\bf H}^{3,8}_{\mu}{\bf H}^{3,8}_{\nu}}$ or 
$$H^3=\frac{|b_1-b_2|}{2r\sin{\vartheta}},\>
H^8=\frac{|b_1+b_2|\sqrt{3}}{2r\sin{\vartheta}}\>,\eqno(39)$$
where we shall put $\sin{\vartheta}=1$ for simplicity.
Table 2 contains estimates of magnetic colour field strength in the ground 
state of charmonium whereas the Bohr radius 
$a_0=0.529177249\cdot10^{5}\ {\rm fm}$ \cite{pdg}. Also when calculating we 
applied the relations $1\ {\rm GeV^{-1}}\approx0.21030893\ {\rm fm}\>$, 
$1\ {\rm T}\approx0.692508\times 10^{-15}\ {\rm GeV}^2$.

\begin{table}[htbp]
\caption{Magnetic colour field strengths in the ground 
state of charmonium.}
\label{t.2}
\begin{center}
\begin{tabular}{|lll|}
\scriptsize $\eta_c(1S)$: & \scriptsize $R_0= 0.0399766\ {\rm fm}$  & \\
\hline
$r$ &  $H^3$ &  $H^8$ \\
 (fm) &  $({\rm T})$ & $({\rm T})$\\
\hline
 $0.1R_0$ &  $0.124857\cdot10^{19}$  &  $0.500516\cdot10^{18}$ \\
\hline         
$R_0$ & $0.124857\cdot10^{18}$  & $0.500516\cdot10^{17}$  \\
\hline   
$10R_0$ & $0.124857\cdot10^{17}$  & $0.500516\cdot10^{16}$  \\
\hline     
$1.0$ & $0.499137\cdot10^{16}$  & $0.200089\cdot10^{16}$  \\
\hline   
$a_0$ & $0.943232\cdot10^{11}$ & $0.378114\cdot10^{11}$  \\
\hline  
\end{tabular}
\end{center}
\end{table}
It is seen that at the characteristic scales of 
the charmonium ground state the strength of magnetic colour field 
responsible for linear confinement reaches huge values of order 
$10^{17}$--$10^{18}\ {\rm T}$. For comparison one should notice that the most 
strong magnetic fields known at present have been discovered in magnetic 
neutron stars, pulsars (see, e. g., Ref. \cite{Bo}) where the corresponding 
strengths can be of order $10^{9}$--$10^{10}\ {\rm T}$. So the characterictic 
feature of confinement is really very strong magnetic colour field between 
quarks. In a certain sense the essence of confinement can be said to be just 
in enormous gluon concentrations and magnetic colour field strentghs in space 
around quarks.

\section{Concluding remarks}
The aim of this Letter we 
pursued was to specify a scenario of linear confinement of quarks, at any rate, 
for mesons and quarkonia. As we could see, crucial 
step consisted in studying exact solutions of the SU(3)-Yang-Mills 
equations modelling confinement. Namely structure of the confining solutions 
for SU(3)-Yang-Mills equations prompts to the idea that linear confinement 
should be governed by magnetic colour field. But the latter is essentially 
relativistic object vanishing in nonrelativistic limit which implies 
the relativistic effects to be extremely important for the confinement 
mechanism \cite{{Gon03},{GC03},{GB04}}.

  Further specification of the confinement mechanism can consist in  
the inquiry answer: what the way of gluon exchange between quarks is? 
The matter concerns such a modification of gluon 
propagator which might correspond to linear confinement at large distances. 
As was remarked in Ref. \cite{Gon03}, the sought modification should be 
carried out on the basis of the exact 
solutions of the SU(3)-Yang-Mills equations modelling quark confinement. 
We hope to consider this modification elsewhere.

There is also an interesting possibility of indirect experimental verification 
of the confinement mechanism under discussion. Really solutions (21) point out 
the confinement phase could be in electrodynamics as well. Though 
there exist no elementary charged particles generating a constant magnetic 
field linear in $r$, the distance from particle, after all, if it could generate 
this elecromagnetic field configuration in laboratory then one might study 
motion of macroscopic charged particles in that field. The confining properties 
of the mentioned field should be displayed at classical level too but the exact 
behaviour of particles in this field requires certain analysis of the corresponding 
classical equations of motion.




\end{document}